\documentclass[11pt]{amsart}
\usepackage{amsmath}
\usepackage{amssymb}
\usepackage{amsfonts}
\usepackage{latexsym}
\usepackage{fancyhdr}     
\usepackage{graphics}
\usepackage{color}
\usepackage{mathrsfs}
\usepackage{latexsym}
\usepackage{mathtools}
\usepackage{enumerate}

\newcommand{\norm}[1]{ \parallel #1 \parallel}
\newcommand{\mb}{\mathbb}
\newcommand{\mc}{\mathcal}
\newcommand{\eul}{\mathfrak}

\newcommand{\M}{\eul M}
\newcommand{\D}{\mc D}
\newcommand{\HH}{\mc H}

\newcommand{\id}{{\mb I}}

\def\*{^*}

\newcommand{\LD}{{\mathcal L}^\dagger(\D)}

\newcommand{\ip}[2]{\langle {#1}|{#2}\rangle}

\newtheorem{defn}{Definition}[section]
\newtheorem{prop}[defn]{Proposition}
\newtheorem{thm}[defn]{Theorem}
\newtheorem{lemma}[defn]{Lemma}

\newtheorem{example}[defn]{Example}
\newtheorem{rem}[defn]{Remark}
\def\x{\relax\ifmmode {\mbox{*}}\else*\fi}
\newcommand{\beex}{\begin{example}$\!\!${\bf }$\;$\rm }
\newcommand{\enex}{ \end{example}}
\newcommand{\berem}{\begin{rem}$\!\!${\bf }$\;$\rm }
\newcommand{\enrem}{ \end{rem}}

\begin{document}
\title{Eigenstates on  $O^\ast$-algebras with an application to ladder operators}%

\author{Fabio Bagarello}
\address{Dipartimento d'Ingegneria, Universit\`a di Palermo, I-90128 Palermo, Italy, and I.N.F.N., Catania, Italy}
\email{fabio.bagarello@unipa.it}
\author{Hiroshi Inoue}
\address{Department of Economics, Kyushu Sangyo University, Fukuoka, Japan}
\email{h-inoue@ip.kyusan-u.ac.jp}
\author{Salvatore Triolo}
\address{Dipartimento d'Ingegneria, Universit\`a di Palermo, I-90128 Palermo, Italy}
\email{salvatore.triolo@unipa.it}

\begin{abstract}
\noindent Following some recent results, we consider here a particular class of states on a  $O^\ast$-algebra $\M$ which we call {\em eigenstates of an element} $A\in\M$, and we deduce some of their properties. Then, we apply our definition to a family of ladder elements, i.e. elements of $\M$ obeying certain algebraic (commutation) relations which are physically motivated, and we discuss several results for these states.
\end{abstract}
\maketitle

\section{Introduction}

The role of eigenstates is quite well known both in pure and in applied mathematics, and it is particularly relevant in quantum mechanics. In this case, for instance, eigenvectors of a given Hamiltonian (which usually describes the energy of a given physical system) are typically interpreted as the stationary states, i.e. those states describing a constant probability.  This is one of the (physical) reasons why the properties of the set of all the eigenstates of a given operator have been studied at length during the past decades. Quite generally, in the literature, eigenstates are meant to be vectors in the Hilbert space, even if some recent results show that eigenvectors of a some particular operators can also be (tempered) distributions, \cite{bagspringer}. On a different side, recently, \cite{denittis}, the authors considered the notion of eigenvectors in a C*-algebraic settings, replacing vectors with linear functional. C*-algebras are often used in connection with quantum systems with infinite degrees of freedom in view of the possibility of using the same underlying structure to describe different physical situations, using inequivalent representations of the abstract C*-algebra describing the system,  \cite{br}-\cite{sewbook2}, which correspond to different physical phases of the system. In this perspective, states are, for C*-algebras, what vectors are for Hilbert spaces. Indeed these are equivalent because of the well known GNS-representation, see \cite{br} for instance.

More recently, in \cite{bit}, we have extended the notion of eigenstates to quasi *-algebras. The reason for this is that  quasi *-algebras are a natural framework to deal with unbounded operators, which appear quite often in many concrete physical systems, see \cite{schm,bag1} and references therein. Our focus in \cite{bit} was on a particular set of elements of a quasi *-algebra which behave as ladder operators and are related to two set of sesquilinear forms for which it is possible to introduce, via generalized GNS, a notion of biorthogonality.  

In this paper we further extend the notion of eigenvectors to  $O^\ast$-algebras. Of course, also this extension can be useful in presence of unbounded operators, for which a C*-algebra is not the most natural settings. In particular,  after some preliminaries, we introduce our definition of eigenstates, and we discuss some results connected with eigenvalues and spectrum of a given element of an O*-algebra $\M$. This is done in Section \ref{sect3}. In Section \ref{sect4} we deduce several properties of eigenstates in connection with a suitable algebraic dynamics. Section \ref{sect5} is devoted to a detailed example based on a class of {\em ladder operators}.

\section{Preliminaries}\label{sect2}
In order to keep the paper self-contained, we collect
below some preliminary definitions and propositions on $O^\ast$-algebras and states on them that
will be used in what follows.

We begin with reminding what an $O^\ast$-algebra is.
Let $\D$ be a dense subspace of a Hilbert space $\HH.$  
We denote by $\mathcal{L}(\D)$ the set of all closable linear operators from  $\D$ to $\D$ and
$\LD$ be the space  consisting of all its elements  which leave, together with their adjoints, the domain $\D$ invariant. Then $\mathcal{L}(\D)$  is an algebra with respect to the usual operations  and   $\LD$ is a subalgebra of  $\mathcal{L}(\D)$ with the involution $X\mapsto X^\dagger := X^\ast\lceil_\D$.
A subalgebra of $\mathcal{L}(\D)$ is called an $O$-algebra on $\D$ in $\HH$, and a $\ast$-subalgebra of $\LD$ is called an $O^\ast$-algebra on $\D$ in $\HH$.
Let $\M$ be an $O^\ast$ algebra on $\D.$
A natural topology on $\D$ defined by a family $\{\norm{\cdot}_X, X\in\M \}$ of seminorms: $$\norm{\xi}_X:=\norm{\xi}+\norm{X\xi}, \quad \xi\in\D$$ is called the graph topology on $\D$ and denoted by $t_{\mathcal{M}}.$
We denote by $\overline{\mathfrak{K}}$ the closure of a subspace $\mathfrak{K}$ of $\HH$ with respect to Hilbert norm topology.
Let $\tilde{\D}[t_{\mathcal{M}}]$ be the completion of the locally convex space equipped with the topology $t_{\mathcal{M}}$ for simplicity denoted $\tilde{\D}({\M}).$
We put $$\tilde{X}:=\overline{X}\upharpoonright_{\tilde{\D}({\M})}, X\in\M, \quad \mbox{and} \quad \tilde{\M}=\{\tilde{X}; X\in\M  \} .$$
Then $\tilde{\M}$ is an $O-$algebra on $\tilde{\D}(\M)$ satisfying $$\tilde{\D}({\M})\subset \hat{D}(\M):=\bigcap_{X\in\M}D(\overline{X}).$$

If $\M$ is an $O^\ast$-algebra on $\D$, then $\tilde{\M}$ is an $O^\ast$-algebra on $\tilde{\D}(\M)$ satisfying $\tilde{\D}(\M)=\hat{\D}(\M)$.
If $\M=\tilde{\M}$, then it is called closed.
We see that $\tilde{\M}$ is the smallest closed extension of $\M$,
so $\tilde{\M}$ is closed the closure of $\M$.
Let $\M$ be an $O^\ast$-algebra on $\D$ in $\HH$.
If $\D^\ast(\M):= \bigcap_{X\in\M}D(X^\ast)=\D$,
then $\M$ is called self-adjoint. This notion will not be used in this paper. However, we prefer to keep it here since it displays well the difference between algebras of bounded and unbounded operators. 
Clearly, if $\M$ is self-adjoint, then it is closed.
Next we define a weak commutant $\M_{\rm w}^\prime$ of $\M$ by
$$ \M_{\rm w}^\prime=\{ C\in B(\HH); (CX\xi|\eta)=(C\xi|X^\dagger \eta)\;\;\; {\rm for \; all}\;\; X\in\M\;\;{\rm and}\;\; \xi,\eta\in\D\},$$
where $B(\HH)$ is a $C^\ast$-algebra of all bounded linear operators on $\HH$.
Then $\M_{\rm w}^\prime$ is a weakly closed $\ast$-preserving subspace of $B(\HH)$,
but it is not necessarily a von Neumann algebra on $\HH$.
If $\M$ is self-adjoint, then $\M_{\rm w}^\prime$ is a von Neumann algebra satisfying $\M_{\rm w}^\prime\D\subset\D$.
We define two locally convex topologies on $\M$.
The topology on $\M$ defined by the family $\{ P_{\xi,\eta}(\cdot); \; \xi,\eta\in\D\}$ (resp. $\{ P_\xi(\cdot); \; \xi\in\D\}$) of seminorms is called the weak (resp. strong) topology and denoted by $\tau_{\rm w}$ (resp. $\tau_{\rm s}$), where
$$P_{\xi,\eta}(X)=|(X\xi|\eta)| \;\;\;{\rm and}\;\;\; P_\xi(X)=\| X\xi\|, \;\; X\in\M, \;\; \xi,\eta\in\D.$$
We refer to \cite{schm}-\cite{powers} for more aspects of $O^\ast$-algebras.
Next we review few aspects of states of an $O^\ast$-algebra $\M$ on $\D$ in $\HH$, focusing in particular in their connected GNS-representations.
Let $\omega$ be a positive linear functional on $\M$.
Then $N_\omega := \{ X\in\M; \;\; \omega(X^\dagger X)=0\}$ is a left ideal of $\M$.
For any $X\in\M$ denote by $\lambda_\omega(X)$ the coset of the quotient space $\M/N_\omega$ which contains $X$,
and define by $(\lambda_\omega(X)|\lambda_\omega(Y))=\omega(Y^\dagger X)$, $X,Y\in\M$ an inner product on $\lambda_\omega(\M)$.
Let $\HH_\omega$ be the Hilbert space which is the completion of the pre-Hilbert space $\lambda_\omega(\M)$.
We define an unbounded $\ast$-representation $\pi_\omega$ of $\M$ in $\HH_\omega$ with domain $\lambda_\omega(\M)$ by $\pi_\omega(A)\lambda_\omega(X)=\lambda_\omega(AX)$ for $A,X\in\M$, that is, $\pi_\omega$ is a $\ast$-homomorphism of the $\ast$-algebra $\M$ into the $\ast$-algebra {$\mathcal{L}^\dagger(\lambda_\omega(\M))$},
and denote the closure of $\pi_\omega$ by the same $\pi_\omega$.
Then $\pi_\omega$ is a closed $\ast$-representation of $\M$ on $\D(\pi_\omega):= \overline{\lambda_\omega(\M)}[t_{\pi_\omega(\M)}]$ in $\HH_\omega$,
and $\xi_\omega:=\lambda_\omega(\id)$ is a strongly cyclic vector for $\pi_\omega(\M)$,
namely $\pi_\omega(\M)\xi_\omega$ is $t_{\pi_\omega(\M)}$-dense in $\D(\pi_\omega)$,
and $(\pi_\omega,\HH_\omega,\lambda_\omega)$ is called the GNS-representation of $\M$.
A positive linear functional $\omega$ on $\M$ with $\omega(\id)=1$ is called a state on $\M$. We denote by $E(\M)$ the set of all states on $\M$.
A state $\omega\in E(\M)$ is called pure if it can not be written in the form $\omega=t\omega_1+(1-t)\omega_2$ with $\omega_1 \neq \omega_2\in E(\M)$ and $0<t<1$.
{By \cite{powers}}  Theorem 6.3, we see that $\omega$ is pure if and only if $\pi_\omega$ is irreducible, that is, $\pi_\omega(\M)_{\rm w}^\prime=\mathbb{C}\id$.
{Concerning the continuity of $\omega\in E(\M)$, if $A\in \M \mapsto \omega(A)\in\mathbb{C}$ is $\tau_\omega$ (resp. $\tau_s$)-continuos, then $\omega$ is called weakly (resp. strongly) continuous. More details, and more useful definitions, will be given later.}\\

\section{Eigenstates and  spectrum}\label{sect3}

\subsection{Eigenstates}

Let $\M$ be an $O^\ast$-algebra on $\D,$ in $\HH$ with identity $\id .$ Following \cite{denittis,bit}, we propose here the following

\begin{defn}\label{defn31}  A state $ \omega\in E(\M) $ is an eiegenstate of $A\in\M$ with eigenvalue $\lambda$ if and only if
	$$\omega(XA)=\lambda\;\omega(X), \quad \forall X\in\M.$$
\end{defn}

Of course this definition implies that, if $A=A^*$, and $\omega$ is its eigenstate with eigenvalue $\lambda$, then $\lambda\in\mathbb{R}$. Indeed we have, taking in particular $X=\id$ and recalling that $\omega(\id)=1$, 
$$
\omega(A)=\lambda\omega(\id)=\lambda \qquad \overline{\lambda}=\overline{\omega(A)}=\omega(A^*)=\omega(A)=\lambda,
$$
as we had to check.

The set of all eigenstates of $A$  will be denoted with $E(A).$
The set of all eigenvalues of $A$ will be denoted with $Eig(A)$.

Let now  $\omega\in E(\M)$ and $(\pi_\omega,\HH_\omega,\lambda_\omega) $ the corresponding cyclic GNS representation. 

\begin{lemma}\label{lemma1} Let $\M$ be  an $O^*-$algebra on $\D,$ in $\HH$ with identity $\id .$ Then $\omega\in E(A)$ with eigenvalue $\lambda$ if and only if $\pi_\omega(A)\xi_\omega=\lambda\;\xi_\omega$, that is, $\lambda$ is an eigenvalue of $\pi_\omega(A)$ with an eigenvector $\xi_\omega$.
\end{lemma}
\begin{proof}
Suppose $\omega\in E(A)$ with eigenvalue $\lambda$.
For every $X\in\M$ we have
\begin{eqnarray*}
<\pi_\omega(A)\xi_\omega|\pi_\omega(X^\dagger)\xi_\omega>
&=& \omega(XA)=\lambda\;\omega(X)\\
&=& <\lambda\xi_\omega|\pi_\omega(X^\dagger)\xi_\omega>.
\end{eqnarray*}
Since $\pi_\omega(\M)\xi_\omega$ is dense in $\HH_\omega$, we conclude that $\pi_\omega(A)\xi_\omega=\lambda\xi_\omega$.

The converse follows from
\begin{eqnarray*}
\omega(XA)
&=& <\pi_\omega(A)\xi_\omega|\pi(X^\dagger)\xi_\omega>\\
&=& \lambda<\xi_\omega|\pi_\omega(X^\dagger)\xi_\omega>\\
&=& \lambda\omega(X)
\end{eqnarray*}
for all $X\in\M$.
\end{proof}

\subsection{Spectrum}

Let $\M$ be an $O^\ast$-algebra on $\D$ in $\HH$ and $A\in\M$.
We define the following notions of spectrum of $A$: 
{$\sigma_p(A):=\{ \lambda\in\mathbb{C}; \;A-\lambda\id$ does not have inverse$\}$} (point spectrum), $Eig(A)$ in Definition 3.1 and
$\sigma_p^\M(A):=\{\lambda\in\mathbb{C}; \; A-\lambda\id$ does not have inverse in $\M\}$. 
For these sets we have the following
\begin{lemma}
Let $A\in\M$. Consider the following statements:
\begin{enumerate}[(i)]
\item $\lambda\in\sigma_p(A)$.
\item There exists a vector $\xi_0$ in $\D$ with $\|\xi_0\|=1$ such that the vector state $\omega_{\xi_0}$ on $\M$ defined by $\omega_{\xi_0}=(X\xi_0|\xi_0)$, $X\in\M$, is an eigenstate of $A$ with an eigenvalue $\lambda$.
\item $\lambda\in Eig(A)$.
\item $\lambda\in\sigma_p^\M(A)$.
\end{enumerate}
Then we have
$$(i)\Rightarrow(ii)\Rightarrow(iii)\Rightarrow(iv)$$
while the converse implications do not necessarily hold.
However, (i) and (ii) are equivalent if and only if $\M\xi_0$ is dense in $\HH$ for some $\xi_0$ in (ii).
\end{lemma}
\begin{proof}
(i)$\Rightarrow$(ii) Let $\lambda\in\sigma_p(A)$.
Then $A\xi_0=\lambda\xi_0$ for some $\xi_0\in\D$ with $\|\xi_0\|=1$.
Hence we have 
$$\omega_{\xi_0}(XA)=(XA\xi_0|\xi_0)=\lambda(X\xi_0|\xi_0)=\lambda\omega_{\xi_0}(X)$$
for all $X\in\M$,
which implies that $\lambda\in Eig(A)$ for $\omega_{\xi_0}\in E(A)$,
and (ii)$\Rightarrow$(i) holds if and only if $\M\xi_0$ is dense in $\HH$.\\
(ii)$\Rightarrow$(iii) This is trivial.
Since any state of $\M$ is not necessarily a vector state, (iii)$\Rightarrow$(ii) does not hold in general.

(iii)$\Rightarrow$(iv) If $\lambda\in Eig(A)$ then by Lemma \ref{lemma1} there exist $\omega\in E(\M)$  such that $\pi_\omega(A)\xi_\omega=\lambda\xi_\omega$, with $\xi_\omega\neq0$. Now we assume that $\lambda\notin\sigma_p^\M(A)$, so that $A-\lambda I$ has inverse in $\M$. Hence
$\xi_\omega=\pi_\omega((A-\lambda\id)^{-1}(A-\lambda\id))\xi_\omega=\pi_\omega((A-\lambda\id)^{-1})(\pi_\omega(A)-\lambda\id)\xi_\omega=0$.
This is a contradiction. Thus $\lambda\in\sigma_p^\M(A)$.
It is clear that (iv)$\Rightarrow$(iii) does not necessarily hold.

\end{proof}

{\bf Remark:--} As often happens for unbounded operators algebras, the notion of spectrum and eigenvalues of their elements not necessarily coincide. The above lemma describes that this is indeed the case in our settings.

\section{Eigenstates and  dynamics}\label{sect4}

\subsection{Dynamics}
In this section let $\M$ be an $O^*-$algebra on $\D\subset\HH$ with identity $\id $ and let $H=H^\dagger\in\M$ 
be essentially self-adjoint.
We remark that $H^\dagger=H$ does not necessarily imply $H$ is essentially self-adjoint differently to bounded case.
Here we denote the self-adjoint extension of $H$ with the same symbol for simplicity.
Then $\{ e^{iHt} \}_{t\in\mathbb{R}}$ is a strongly continuous one-parameter unitary group on  $\HH.$

The main assumption in this section is the following:\\

[A1]: $ e^{itH}\in\M \quad \mbox{for all } t\in[0,1]$. \\

This implies that $ e^{itH}\in\M$ for all  $t\in\mathbb{R}$. In fact, given $t\in\mathbb{R}$, is surely exists $k\in\mathbb{N}$ such that $\frac{t}{k}=:t_k\in[0,1]$. Hence, since $e^{itH}=(e^{it_KH})^k$, our claim follows from the fact that $\M$ is an algebra.

\vspace{2mm}

{\bf Remark:--} This simple result has an interesting consequence, based on what follows: a {\em local} (in time) dynamics can be extended to a {\em global} (again, in time) dynamics. This is quite important for concrete physical systems.

\vspace{2mm}

Under this assumption, if we put $\alpha_t^{H}(A):= e^{itH}A e^{-itH}, \quad A\in\M, \, t\in\mathbb{R}.$
Then $\{\alpha_t^{H}(A)\}_{t\in\mathbb{R}}$ is a one-parameter group of *-automorphisms of $\M$ and 
$(\M,\{\alpha_t^{H} \} )$ is what we can call {\em an  (algebraic) dynamical system}.

We recall that the map $\{\alpha_t^H\}_{t\in\mathbb{R}}$ is strongly  continuous if for all $A\in\M$ $\lim_{t\rightarrow 0}\|\alpha_t^H(A)\xi-A\xi\|=0$ for all $\xi\in\D$. Also, $\{\alpha_t^H\}_{t\in\mathbb{R}}$ is weakly continuous if for all $A\in\M$, $\lim_{t\rightarrow0} |<\alpha_t^H(A)\xi|\eta>-<A\xi|\eta>|=0$ for all $\xi,\eta\in\D$.
 Lastly,
	$\{e^{itH} \}_{t\in\mathbb{R}}$ is $ t_{\M}-$ continuous if for any $\xi\in\D$ $\lim_{t\rightarrow0}\| Ae^{itH}\xi-A\xi\|=0$ for all $A\in\M$.

It is possible to prove the following

\begin{lemma}
Under Assumption [A1] the following statements are equivalent:

[i] $\{\alpha_t^{H} \}_{t\in\mathbb{R}}$  is strongly continuous.

[ii] $ \{\alpha_t^{H} \}_{t\in\mathbb{R}}  $ is weakly continuous.

[A2] $ \{e^{itH} \}_{t\in\mathbb{R}}$ is $ t_{\M}-$ continuous.

\end{lemma}
\begin{proof}
$(i)\Rightarrow[A2]:$ For every $A\in\M$ and  $\xi \in \D$ we have
\begin{eqnarray*}    \norm{Ae^{itH}\xi-A\xi}&=&\norm{e^{-itH}Ae^{itH}\xi-e^{-itH}A\xi}=\norm{(e^{-itH}Ae^{itH}\xi-A\xi)+(A\xi-e^{-itH}A\xi)}  \\&\leq&\norm{\alpha^H_{-t}(A)\xi-A\xi}+ \norm{A\xi -e^{-itH}A\xi}\rightarrow 0\;\;\;{\rm as}\;\; t\rightarrow 0,\end{eqnarray*}
where we have also used the fact that $e^{-itH}$ is strongly convergent to the identity operator of $\M$, for $t\rightarrow0$.

$[A2]\Rightarrow(i):$ For every $A\in\M$
		.\begin{eqnarray*}  
	 \norm{\alpha^H_{t}(A)\xi-A\xi}&\leq&\norm{e^{itH}Ae^{-itH}\xi-e^{itH}A\xi}+\norm{ e^{itH}A\xi-A\xi}  \\&=&\norm{Ae^{-itH}\xi-A\xi}+ \norm{e^{itH}A\xi -A\xi}\rightarrow 0\;\;\;{\rm as}\;\; t\rightarrow 0.		
\end{eqnarray*}
$(i)\Rightarrow(ii):$ It is trivial.\\
$(ii)\Rightarrow(i):$  For every $A\in\M$
.\begin{eqnarray*}  
\norm{\alpha^H_{t}(A)\xi-A\xi}^2&=&\norm{e^{itH}Ae^{-itH}\xi}^2-\ip{e^{itH}A\xi}{A\xi}-\ip{A\xi}{e^{itH}A\xi}+\norm{A\xi}^2  \\&\rightarrow& 0\;\;\;{\rm as}\;\; t\rightarrow 0,	
\end{eqnarray*}
 because
 
 \begin{eqnarray*}  
 	\norm{e^{itH}Ae^{-itH}\xi}^2 &=&\norm{Ae^{-itH}\xi}=\ip{Ae^{-itH}\xi}{Ae^{-itH}\xi}\\&=&\ip{e^{itH}A^\dagger Ae^{-itH}\xi}{\xi}=\ip{\alpha_t^{H}(A^\dagger A)\xi}{\xi}.\\
 	 		 \end{eqnarray*}
 Therefore $\lim_{t\rightarrow0}\norm{\alpha^H_{t}(A)\xi-A\xi}=0.$ 
\end{proof}

\vspace{2mm}

{\bf Remark.} It is useful to emphasize that Assumption [A1] does not necessarily imply Assumption [A2]: given $\xi\in\D$ such that $\lim_{t\rightarrow0}\|e^{itH}\xi-\xi\|=0$, it is known that
for a generic unbounded operator $A\in\M$ $\lim_{t\rightarrow0}\|Ae^{itH}\xi-A\xi\|=0$ does not necessarily hold. However, for any $A\in\M$ $Ae^{itH}\xrightarrow[t\rightarrow0]{} A$, weakly, that is, for any $\xi,\eta\in\D$ $<Ae^{itH}\xi|\eta>=<e^{itH}\xi|A^\dagger\eta>\rightarrow<\xi|A^\dagger\eta>=<A\xi|\eta>$ holds.

\vspace{2mm}

From now on, in this Section 4.1, we assume both [A1] and [A2]. Then we have the following 

{\begin{lemma}\label{lemma2}
For every $A\in \M$ we have $\lim_{t\rightarrow0} \; \frac{\alpha_t^{H}(A)-A}{t}=i[H,A]\xi=:\delta_H(A),$ weakly.
\end{lemma}
\begin{proof}
By assumption [A2] and Lemma 4.1 we have
\begin{eqnarray*}
\lim_{t\rightarrow\infty} \frac{e^{itH}-I}{t}\xi
&=&\lim_{t\rightarrow\infty}iHe^{itH}\xi \\
&=& iH(\lim_{t\rightarrow\infty} e^{itH})\xi \\
&=& iH\xi
\end{eqnarray*}
for all $\xi\in\D$, which implies that for any $\xi, \eta\in\D$
\begin{eqnarray*}
\lim_{t\rightarrow\infty}<\frac{\alpha_t^H(A)-A}{t}\xi|\eta>
&=& \lim_{t\rightarrow\infty} \left( <\frac{e^{itH}Ae^{-itH}-e^{itH}A}{t}\xi|\eta> +<\frac{e^{itH}-I}{t}A\xi|\eta>\right)\\
&=& \lim_{t\rightarrow\infty} \left( <\frac{e^{-itH}-I}{t}\xi|A^\dagger e^{-itH}\eta>+<\frac{e^{itH}-I}{t}A\xi|\eta>\right)\\
&=& <-iH\xi|A^\dagger\eta>+<iHA\xi|\eta>\\
&=& <-iAH\xi+iHA\xi|\eta>\\
&=& <i(HA-AH)\xi|\eta>.
\end{eqnarray*}
Hence $\frac{\alpha_t^H(A)-A}{t}\xrightarrow[t\rightarrow0]{} \delta_H(A)$, weakly.
\end{proof}}
\begin{defn}
Given a state $\omega\in E(\M)$ we say that $\omega$ is invariant under $\alpha^H$ if $\omega(\alpha_t^H(A))=\omega(A)$ for every $A\in \M$ and $t\in\mathbb{R}.$
\end{defn}

It is easy to prove the following Theorem, which extends to O*-algebras some well known results in C*-algebras:
\begin{thm} 
Let $\omega\in E(\M)$. Then we have the following 

(i) If {$\omega$ is weakly continuous and it }is invariant under   $\alpha^H$ then $\omega(\delta_H(A))=\omega(i[H,A])=0$  for every $A\in \M$ and $t\in\mathbb{R}.$ 

(ii) If $\omega$ is an eigenstate of $H$ with eigenvalue $\lambda$ then $\omega$ is an eigenstate $H^n$ with eigenvalue $\lambda^n$, $n\geq0$. Moreover, $\omega$ is also an eigenstate of $e^{itH}$ with eigenvalue $e^{it\lambda}.$

(iii) If $\omega$ is an eigenstate of $H$ with eigenvalue $\lambda\in\sigma(H)$ then $\omega$ is invariant under $\alpha^{H}.$

(iv) If $\omega$ is a weakly continuous pure state invariant under   $\alpha_t^H$ then $\omega$ is an eigenstate of $H.$  

\end{thm}
\begin{proof}
$(i)$ Since $\omega$ is {weakly} continuous, we have $$\omega(\delta_H(A))=\lim_{t\rightarrow0}\omega(s-\frac{\alpha_t^{H}(A)-A}{t})=\lim_{t\rightarrow0}\frac{\omega(\alpha_t^{H}(A)-A)}{t}=\lim_{t\rightarrow0}\frac{1}{t}\left(\omega(\alpha_t^{H}(A)-A)\right)=0,$$
since $\omega(\alpha_t^{H}(A)=\omega(A)$.

$(ii)$ First we observe that, since $\omega(AH)=\lambda\omega(H)$ for all $A\in \M$, then, for instance, $\omega(AH^2)=\omega((AH)H)=\lambda\omega(AH)=\lambda^2\omega(A)$, for all $A\in\M$. Higher powers work in the same way.
To prove the second statement, we first notice that, for all $A\in\M$ and for all $n\geq0$, 
$$
\omega(H^nA)=\lambda^n\omega(A).
$$
Indeed we have that, using the fact that $H=H^\dagger$, our claim above and the reality of $\lambda$,
$$
\omega(H^nA)=\overline{\omega(A^\dagger H^n)}=\lambda^n\overline{\omega(A^\dagger)}=\lambda^n\omega(A).
$$
Notice now that, for every $A\in\M$ and $\xi\in\D$ we have $e^{itH}A\xi=\lim_{n\rightarrow \infty} \sum_{h=0}^{n}\frac{(itH)^h}{h!}A\xi$.
Since $\omega$ is strongly continuous, we have
\begin{eqnarray*}  \omega(e^{itH}A)&=&\sum_{n=0}^{\infty}\frac{(it)^n\omega(H^nA)}{n!}
		\\&=&\sum_{n=0}^{\infty}\frac{(it)^n}{n!}\lambda^n\omega(A)=e^{it\lambda}\omega(A),\end{eqnarray*}
	so that
\begin{eqnarray*} 		
	\omega(Ae^{itH})
	&=& \overline{\omega((Ae^{itH})^\dagger)}\\
	&=& \overline{\omega(e^{-itH}A^\dagger)}\\
	&=& \overline{e^{-it\lambda}\omega(A^\dagger)} \\
	&=& e^{it\lambda}\omega(A).
\end{eqnarray*}
(iii) For every $A\in\M,$ by (ii) 	$\omega(Ae^{itH})=e^{it\lambda}\omega(A).$ Moreover 
\begin{eqnarray*}  
\omega(\alpha_t^{H}(A))&=&\omega(e^{itH}Ae^{-itH})\\&=&e^{-itH}\omega(e^{itH}A)=\omega(A).
\end{eqnarray*}
(iv) Since $\omega$ is $\alpha_t^H$-invariant, we have
$$\| \pi_\omega(\alpha_t^H(A))\xi_\omega\|=\|\pi_\omega(A)\xi_\omega\|$$
for all $A\in\M$ and $t\in\mathbb{R}$.
Hence we can define a strongly continuous one-parameter unitary group $\{ U_\omega(t)\}$ on $\HH_\omega$ by $U_\omega(t)\pi_\omega(A)\xi_\omega=\pi_\omega(\alpha_t^H(A))\xi_\omega$, $A\in\M$, $t\in\mathbb{R}$.
By Stone's theorem there exists a self-adjoint operator $K_\omega$ on $\HH_\omega$ such that $U_\omega(t)=e^{itK_\omega}$ for all $t\in\mathbb{R}$.
Since
$$\xi_\omega=U_\omega(t)\xi_\omega=e^{itK_\omega}\xi_\omega= \sum_{n=0}^\infty \frac{(itK_\omega)^n}{n!}\xi_\omega,$$
we have $K_\omega\xi_\omega=0$.
Moreover, we see
\begin{equation} 
\pi_\omega(\alpha_t^{H}(A))=\pi_\omega(e^{itH})\pi_\omega(A)\pi_\omega(e^{-itH})=e^{it K_\omega}\pi_\omega(A)e^{-itK_\omega} \nonumber
\end{equation}
for all $A\in\M$ and $t\in\mathbb{R}$.
Indeed, for any $B\in\M$
\begin{eqnarray*}
\pi_\omega(A)\pi_\omega(e^{-itH})e^{itK_\omega}\pi_\omega(B)\xi_\omega
&=& \pi_\omega(A)\pi_\omega(e^{-itH})\pi_\omega(\alpha_t^H(B)) \xi_\omega\\
&=& \pi_\omega(A)\pi_\omega(Be^{-itH})\xi_\omega\\
&=& \pi_\omega(e^{-itH}\alpha_t^H(AB))\xi_\omega\\
&=& \pi_\omega(e^{-itH})U_\omega(t)\pi_\omega(AB)\xi_\omega\\
&=& \pi_\omega(e^{-itH})e^{itK_\omega}\pi_\omega(A)\pi_\omega(B)\xi_\omega,
\end{eqnarray*}
which implies $\pi_\omega(e^{-itH})e^{itK_\omega}\in\pi_\omega(\M)_{\rm w}^\prime=\mathbb{C}\id$
since $\lambda_\omega(\M)$ is $t_{\pi_\omega(\M)}$-dense in $\D(\pi_\omega)$.
Hence $e^{itK_\omega}=\gamma_t\pi_\omega(e^{itH})$ for some $\gamma_t\in\mathbb{C}$ with $|\gamma_t|=1$, which implies that
\begin{equation} \label{c}
\pi_\omega(\alpha_t^{H}(A))=\pi_\omega(e^{itH})\pi_\omega(A)\pi_\omega(e^{-itH})=e^{it K_\omega}\pi_\omega(A)e^{-itK_\omega} 
\end{equation}
for all $A\in\M$ and $t\in\mathbb{R}$.
For every $X,Y\in\M,$ by Lemma 4.2 we have
\begin{equation} 
\lim_{t\rightarrow0}\ip{Y^\dagger(\frac{\alpha_t^{H}(A)-A}{t})X\xi}{\eta}=\lim_{t\rightarrow0}\ip{\frac{\alpha_t^{H}(A)-A}{t}X\xi}{Y\eta}=\ip{i[H,A]X\xi}{Y\eta}\;\;{\rm as}\;\; t\rightarrow 0 \nonumber
\end{equation}
for every $\xi,\eta\in\D$.
Since $\omega$ is weakly continuous,
$$\lim_{t\rightarrow 0} \omega({Y^\dagger(\frac{\alpha_t^{H}(A)-A}{t})X-iY^\dagger[H,A]X})=0.$$
Thus 
\begin{eqnarray*} 
	\lim_{t\rightarrow 0}\ip{\frac{\pi_\omega(\alpha_t^{H}(A))-\pi_\omega(A)}{t}-i[\pi_\omega(H),\pi_\omega(A)]\pi_\omega(X)\xi_\omega}{\pi_\omega(Y)\xi_\omega}=0,
	\end{eqnarray*}
and 
\begin{eqnarray*} 
	\lim_{t\rightarrow 0}\ip{\frac{\pi_\omega(\alpha_t^{H}(A))-\pi_\omega(A)}{t}\pi_\omega(X)\xi_\omega}{\pi_\omega(Y)\xi_\omega}=i\ip{[\pi_\omega(H),\pi_\omega(A)]\pi_\omega(X)\xi_\omega}{\pi_\omega(Y)\xi_\omega}
\end{eqnarray*}
Thus by \eqref{c} 
\begin{eqnarray*} 
	\lim_{t\rightarrow 0}\ip{\frac{\pi_\omega(\alpha_t^{H}(A))-\pi_\omega(A)}{t}\pi_\omega(X)\xi_\omega}{\pi_\omega(Y)\xi_\omega}&=&i\ip{[\pi_\omega(H),\pi_\omega(A)]\pi_\omega(X)\xi_\omega}{\pi_\omega(Y)\xi_\omega}\\&=&
	\lim_{t\rightarrow 0}\ip{\frac{\alpha_t^{K_\omega}(\pi_\omega(A))-\pi_\omega(A)}{t}\pi_\omega(X)\xi_\omega}{\pi_\omega(Y)\xi_\omega}\\&=&i\ip{[K_\omega,\pi_\omega(A)]\pi_\omega(X)\xi_\omega}{\pi_\omega(Y)\xi_\omega}.
\end{eqnarray*}
Therefore for every $X,Y\in\M$ $$\ip{([K_\omega,\pi_\omega(A)]-[\pi_\omega(H),\pi_\omega(A)])\pi_\omega(X)\xi_\omega}{\pi_\omega(Y)\xi_\omega}=0,$$ since $\pi_\omega(\M)\xi_\omega$ is dense in $\HH_\omega,$ $$[K_\omega,\pi_\omega(A)]=K_\omega\pi_\omega(A)-\pi_\omega(A)K_\omega=\pi_\omega(H)\pi_\omega(A)-\pi_\omega(A)\pi_\omega(H)=[\pi_\omega(H),\pi_\omega(A)].$$
Then for every $A\in\M$ $(K_\omega-\pi_\omega(H))\pi_\omega(A)=\pi_\omega(A)(K_\omega-\pi_\omega(H)).$
Therefore $(K_\omega-\pi_\omega(H)\in\pi_\omega(\M)'_\omega=\mathbb{C}\id.$ Then there exists $\lambda$ such that $\pi_\omega(H)=K_\omega+\lambda.$
Hence $\pi_\omega(H)\xi_\omega=K_\omega\xi_\omega+\lambda\xi_\omega=\lambda\xi_\omega$,
which means that $\omega$ is an eigenstate for $H$ with eigenvalue $\lambda$.
\end{proof}

\subsection{Ground States}
In this section, let $\M$ be an $O^\ast$-algebra on $\D$ in $\HH$ with identity $\id$ and $H^\dagger=H\in\M$.
\begin{defn}
We say that $\omega\in E(\M)$ is a ground state for $H$ if $\overline{\pi_\omega(H)}$ is self-adjoint  and $\omega (H)$ is the minimum of $\sigma(\overline{\pi_\omega(\HH)})$, and denote it by  $\lambda_*$.
\end{defn}

\begin{thm} \label{theorem1}
Let $\omega\in E(\M)$. The following statements are equivalent.

(i)  $\omega$ is a ground state for $H.$

(ii)  $\pi_\omega(H)$ is essentially self- adjoint, $\omega$ an eigenstate of $H$ with $\lambda_*$ and $-i\omega(A^\dagger\delta_H(A))\ge 0$ for all $A\in\M$.
\end{thm}

\begin{proof}
$(i)\Rightarrow(ii):$	We show $\overline{\pi_\omega(H)}\xi_\omega=\lambda_*\xi_\omega.$
	We put, by spectral decomposition of self-adjoint operator 
	\begin{equation} \label{e}
		\overline{\pi_\omega(H)}=\int_{\sigma(\overline{\pi_\omega(H)})}\lambda d_{(\overline{\pi_\omega(H)})}(\lambda).\end{equation}
	If  $\sigma(\overline{\pi_\omega(H)})\setminus\{\lambda_*\}=\emptyset,$ then by \eqref{e}
	$\overline{\pi_\omega(H)}\xi_\omega=\lambda_*\xi_\omega.$
	Then we assume  $\sigma(\overline{\pi_\omega(H)})\setminus\{\lambda_*\}\neq\emptyset.$
	Thus \begin{eqnarray*} 0=\omega(H)-\lambda_*&=&\ip{(\overline{\pi_\omega(H)}-\lambda_*{\id})\xi_\omega}{\xi_\omega}\\&=&\int_{\sigma(\overline{\pi_\omega(H)})}(\lambda-\lambda_*)d_{\ip{E_H(\lambda)\xi_\omega}{\xi_\omega}}\\
	&=& \int_{\sigma(\overline{\pi_\omega(H)})\backslash\{\lambda_\ast\}}(\lambda-\lambda_\ast) d_{\ip{E_H(\lambda)\xi_\omega}{\xi_\omega}}.
\end{eqnarray*} 
Since $\lambda-\lambda_*>0$ on $\overline{\pi_\omega(H)}-\lambda_*,$ we have $\ip{E_H(\sigma(\overline{\pi_\omega(H)})\setminus\{\lambda_*\})}{\xi_\omega}=0.$
Therefore  $$\ip{E_H(\lambda)\xi_\omega}{\xi_\omega}=\ip{E_H(\sigma({\pi_\omega(H)})\xi_\omega}{\xi_\omega}=\omega(\id)=1.$$
Then we have
\begin{eqnarray*}
\norm{E_H(\lambda_*)\xi_\omega-\xi_\omega}^2&=& \norm{E_H(\lambda_*)\xi_\omega}^2-\ip{E_H(\{\lambda_*\})\xi_\omega}{\xi_\omega}\\&-&\ip{\xi_\omega}{E_H(\{\lambda_*\})\xi_\omega}+\norm{\xi_\omega}^2=0.
\end{eqnarray*}
Then $E_H(\lambda_*)\xi_\omega=\xi_\omega,$ thus $\pi_\omega(H)E_H(\lambda_*)\xi_\omega=\lambda_*\xi_\omega=\pi_\omega(H)\xi_\omega.$
Moreover we have
\begin{eqnarray*}
-i\omega(A^\dagger \delta_H(A))&:=&-i\omega(A^\dagger i[H,A])\\
&=&\omega(A^\dagger H A)-\omega((A^\dagger A)H)\\
&=& \omega(A^\dagger HA)-\lambda_*\omega(A^\dagger A)\\&=& \omega(A^\dagger(H-\lambda_*H)A)\\&=&\ip{\pi_\omega(H-\lambda_*\id)\pi_\omega(A)\xi_\omega}{\pi_\omega(A)\xi_\omega}\ge 0
\end{eqnarray*}
for all $A\in\M$.\\
$(ii)\Rightarrow(i):$ Let $\lambda$ be an eigenvalue of an eigenstate $\omega$ for $H$ then $\omega(AH)=\lambda\;\omega(A)$  for every $A\in\M.$ Then, since $-i\omega(A^\dagger\delta_H(A))=\omega(A^\dagger HA)-\omega(A^\dagger A H)\ge 0$  for $A\in\M$,  we have $$\ip{\overline{\pi_\omega(H)} \pi_\omega(A)\xi_\omega}{\pi_\omega(A)\xi_\omega}=\omega(A^\dagger HA)\ge\omega(A^\dagger A H)=\lambda\;\omega(A^\dagger A)=\lambda\norm{\lambda\omega(A)}^2.$$
 Thus $$\lambda\leq \inf_{\norm{\pi_\omega(A)\xi_\omega}=1}\ip{\overline{\pi_\omega(H)} \pi_\omega(A)\xi_\omega}{\pi_\omega(A)\xi_\omega}:=M_{\pi_\omega(H)}.$$
 For every $\mu< M_{\pi_\omega(H)}$ there exists $\epsilon_o>0$ such that $ \mu+\epsilon_o<M_{\pi_\omega(H)}.$
 
 Therefore for every $A\in\M$ with $\norm{\pi_\omega(A)\xi_\omega}=1$ we have  $\ip{\overline{\pi_\omega(H)} \pi_\omega(A)\xi_\omega}{\pi_\omega(A)\xi_\omega}>\mu+\epsilon_o.$
 Since $\pi_\omega(\M)\xi_\omega$ is a core for $\pi_\omega(H),$ then 
 $\ip{\overline{\pi_\omega(H)\xi}}{\xi}\ge (\mu+\epsilon_o)\norm{\xi}^2$ for every $\xi\in\D(\overline{\pi_\omega(H)})$ with $\norm{\xi}=1.$
 Then by the Schwarz inequality  for every $\xi\in\D((\overline{\pi_\omega(H)})$
$$\norm{(\overline{\pi_\omega(H)}-\mu \id)\xi}\norm{\xi}\ge \ip{\overline{(\pi_\omega(H)}-\mu\id)\xi}{\xi}\ge \epsilon_o \norm{\xi}.$$
Then $(\overline{\pi_\omega(H)}-\mu\id)$ has bounded inverse thus $\mu\notin\sigma(\overline{\pi_\omega(H)}).$
On the other hand, since $\overline{\pi_\omega(H)}\xi_\omega=\lambda \;\xi_\omega, $ we have $ \lambda\in\sigma(\overline{\pi_\omega(H)}).$ Thus $\lambda=\lambda_*=\min \sigma(\overline{\pi_\omega(H)}),$ therefore $\omega$ is a ground state for $H.$
\end{proof}

\subsection{Gapped ground states}
Let $\M$  be  $O^*-$algebra on $\D,$ in $\HH$ with identity $\id $ and $H=H^\dagger\in\M.$
\begin{defn}
	A ground state $\omega$ for $H$ is called gapped  if $\overline{\pi_\omega(H)}$ is self-adjoint for some $\Delta>0$ and $$\sigma(\overline{\pi_\omega(H)})\cup(\lambda_*,\lambda_*+\Delta)=\emptyset.$$
\end{defn}

Furthermore, a gapped ground state $\omega$ is called nondegenerate if $Ker(\overline{\pi_\omega(H)}-\lambda_*\id)=\mathbb{C}\xi_\omega.$

The following statements can be proved
\begin{thm}
	Let $\omega$ be a ground state for $H$. The following statements are equivalent.
	
	(i) $\omega $ is gapped.
	
	(ii) $\ip{\overline{\pi_\omega(H)}\eta}{\eta} \geq (\lambda_\ast+\Delta)\| \eta\|^2$ for all $\eta\in Ker(\overline{\pi_\omega(H)}-\lambda_\ast\id)^\perp$ .
	
	(iii) $\omega$ is gapped and nondegenerate.
	
	(iv) There exists $\Delta>0$ such that
	$$ -i\omega(A^\dagger \delta_H(A)) \geq \Delta(\omega(A^\dagger A)-|\omega(A)|^2$$
	for all $A\in\M$.\\
	Then,
	\begin{equation}
\begin{array}{cccc}
(iii)& \Leftrightarrow & (iv) \\
 &\Downarrow&\\
(i)& \Leftrightarrow&(ii).\\
\end{array}
\nonumber
\end{equation}

	\end{thm}
	\begin{proof}
	(i)$\Rightarrow$(ii) Since $\omega$ is gapped,
	$$\overline{\pi_\omega(H)}=\lambda_\ast E_H(\{\lambda_\ast\})+\int_{\lambda_\ast+\Delta}^\infty \lambda d E_H(\lambda),$$
	so $Ker(\overline{\pi_\omega(H)}-\lambda_\ast\id)=E_H(\{\lambda_\ast\})\HH_\omega$.
	Hence, for each $x\in\HH_\omega$ we have
	\begin{eqnarray*}
	<\overline{\pi_\omega(H)}(\id-E_H(\{\lambda_\ast\})x|(\id-E_H(\{\lambda_\ast\}))x>
	&=& <\int_{\lambda_\ast+\Delta}^\infty \lambda dE_H(\lambda)(\id-E_H(\{\lambda_\ast\}))x|(\id-E_H(\{\lambda_\ast\}))x>\\
	&\geq& (\lambda_\ast+\Delta)\|(\id-E_H(\{\lambda_\ast\}))x\|^2.
	\end{eqnarray*}
	(ii)$\Rightarrow$(i) Take an arbitrary $\mu$ with $\lambda_\ast<\mu<\lambda_\ast+\Delta$ and put $\mu=\lambda_\ast+\delta$.
	For any $x\in D(\overline{\pi_\omega(H)})$ we have
	\begin{eqnarray*}
	(\Delta-\delta)\|(\id-E_H(\{\lambda_\ast\})x\|^2
	&\leq& ((\overline{\pi_\omega(H)}-\mu\id)(\id-E_H(\{\lambda_\ast\}))x|(\id-E_H(\{\lambda_\ast\}))x)\\
	&\leq& \|(\overline{\pi_\omega(H)}-\mu\id)(\id-E_H(\{\lambda_\ast\}))x\|\|(\id-E_H(\{\lambda_\ast\}))x\|,
	\end{eqnarray*}
	so
	$$\|(\overline{\pi_\omega(H)}-\mu\id)(\id-E_H(\{\lambda_\ast\})x\| \geq (\Delta-\delta)\|(\id-E_H(\{\lambda_\ast\}))x\|.$$
	Furthermore,
	$$\|(\overline{\pi_\omega(H)}-\mu\id)(E_H(\{\lambda_\ast\}))x\|=\delta \|E_H(\{\lambda_\ast\})x\|.$$
	Hence
	$$\|(\overline{\pi_\omega(H)}-\mu\id)x\| \geq \min(\Delta-\delta)\|x\|$$
	for all $x\in D(\overline{\pi_\omega(H)})$,
	so $(\overline{\pi_\omega(H)}-\mu\id)$ has bounded inverse.
	Thus $\mu \notin \sigma(\overline{\pi_\omega(H)})$, and so $\sigma(\overline{\pi_\omega(H)}) \cap (\lambda_\ast,\lambda_\ast+\Delta)=\emptyset$.
	Thus (i) and (ii) are equivalent.\\
	Next we show (iii) and (iv) are equivalent.
	Take an arbitrary $A\in\M$.
	Then
	\begin{eqnarray*}
	\Omega_A
	&:=& \pi_\omega(A)\xi_\omega-<\xi_\omega\otimes\overline{\pi_\omega(A^\dagger)\xi_\omega}>\xi_\omega\\
	&=& \pi_\omega(A)\xi_\omega-\ip{\xi_\omega}{\pi_\omega(A^\dagger)\xi_\omega}\xi_\omega\\
	&=& \pi_\omega(A)\xi_\omega-\omega(A)\xi_\omega
	\end{eqnarray*}
	and
	\begin{equation} \label{f}
	\ip{\Omega_A}{\Omega_A}
	=(\pi_\omega(A)\xi_\omega-\omega(A)\xi_\omega |\pi_\omega(A)\xi_\omega-\omega(A)\xi_\omega)
	=\omega(A^\dagger A)-|\omega(A)|^2.
	\end{equation}
	Furthermore, since $\omega$ is a ground state of $H$, we have
	\begin{eqnarray*}
	\ip{\overline{\pi_\omega(H)}\Omega_A}{\Omega_A}
	&=& \ip{\overline{\pi_\omega(H)}(\pi_\omega(A)\xi_\omega-\omega(A)\xi_\omega)}{\pi_\omega(A)\xi_\omega-\omega(A)\xi_\omega}\\
	&=&\ip{\pi_\omega(HA)\xi_\omega}{\pi_\omega(A)\xi_\omega}-\overline{\omega(A)}\ip{\pi_\omega(HA)\xi_\omega}{\xi_\omega}\\
	&& -\omega(A)\ip{\pi_\omega(H)\xi_\omega}{\pi_\omega(A)\xi_\omega}+|\omega(A)|^2\ip{\pi_\omega(H)\xi_\omega}{\xi_\omega}\\
	&=& \omega(A^\dagger HA)-\overline{\omega(A)}\omega(HA)-\omega(A)\omega(A^\dagger H)+|\omega(A)|^2 \omega(H)\\
	&=& \omega(A^\dagger HA)-\lambda_\ast |\omega(A)|^2-\lambda_\ast|\omega(A)|^2+\lambda_\ast|\omega(A)|^2\\
	&=& \omega(A^\dagger HA)-\lambda_\ast|\omega(A)|^2,
	\end{eqnarray*}
	so that by \eqref{f}
	\begin{equation} \label{g}
	\begin{split}
	&\ip{\overline{\pi_\omega(H)}\Omega_A}{\Omega_A}-\lambda_\ast \ip{\Omega_A}{\Omega_A}\\
	&= \omega(A^\dagger HA)-\lambda_\ast|\omega(A)|^2-\lambda_\ast(\omega(A^\dagger A)-|\omega(A)|^2)\\
	&=\omega(A^\dagger HA)-\lambda_\ast\omega(A^\dagger A).
	\end{split}
	\end{equation}
	(iii)$\Rightarrow$(iv) Since $\omega$ is nondegenerate, we have
	\begin{eqnarray*}
	\Omega_A
	&\in& \{ \xi_\omega\}^\perp \\
	&=& ( Ker(\overline{\pi_\omega(H)}-\lambda_\ast\id)^\perp\\
	&=& (\id-E_H(\{\lambda_\ast\}))\HH_\omega,
	\end{eqnarray*}
	so by Lemma \ref{lemma1},
	$$\ip{(\overline{\pi_\omega(H)}-(\lambda_\ast+\Delta)\id)\Omega_A}{\Omega_A}\geq 0.$$
	Hence it follows from \eqref{f} and \eqref{g} that
	\begin{eqnarray*}
	-i\omega(A^\dagger \delta_H(A))
	&=&\omega(A^\dagger HA)-\omega(A^\dagger AH)\\
	&=& \omega(A^\dagger HA)-\lambda_\ast\omega(A^\dagger A)\\
	&=& \ip{\overline{\pi_\omega(H)}\Omega_A}{\Omega_A}-\lambda_\ast\|\Omega_A\|^2\\
	&\geq& (\lambda_\ast+\Delta)\|\Omega_A\|^2-\lambda_\ast\|\Omega_A\|^2\\
	&\geq& \Delta\|\Omega_A\|\\
	&=& \Delta(\omega(A^\dagger A)-|\omega(A)|^2).
	\end{eqnarray*}
	(iv)$\Rightarrow$(iii) This follows from Theorem \ref{theorem1}. Hence the proof is completed.
	\end{proof}

\section{An example from quantum mechanics: eigenstates for ladder operators}\label{sect5}

In this section we will describe an example motivated by ladder operators, widely used in quantum mechanics in different problems, from the quantum oscillator to quantum field theory and quantum optics. We refer to \cite{bagspringer,bag2024,bit} for some results in this topic, but in a slightly different algebraic and functional context.

We start noticing that if $\omega\in E(\M)$, then $\omega(x^\dagger)=\overline{\omega(x)}$, $x\in\M$. This is because of the following rather general fact, which we include here for reader's convenience:

 If $f$ is positive linear functional on a $\ast$-algebra $\mathscr{A}$, then $f(x^\ast y)=\overline{f(y^\ast x)}$ for all $x,y\in\mathscr{A}$.

Indeed we have, for any $x,y\in\mathscr{A}$ and $\lambda\in\mathbb{C}$,
\begin{eqnarray*}
	0 \leqq f((x+\lambda y)^\ast(x+\lambda y))
	=f(x^\ast x)+\lambda f(x^\ast y)+\bar{\lambda}f(y^\ast x)+|\lambda|^2 f(y^\ast y).
\end{eqnarray*}
Hence we have $\lambda f(x^\ast y)+\bar{\lambda}f(y^\ast x)\in\mathbb{R}$.
If we choose $\lambda=1$ and $\lambda=i$ we deduce that both $f(x^\ast y)+f(y^\ast x)$ and
$i(f(x^\ast y)-f(y^\ast x))$ must be real. Then we easily conclude that
$f(x^\ast y)=\overline{f(y^\ast x)}$.
 In particular, if $\id \in\mathscr{A}$, then we put $y=\id$ so $f(x^\ast)=\overline{f(x)}$.

Let us now take an element $A\in\M$ with $\omega(AA^\dagger)\neq 0$.
Then $\omega_A:=\frac{1}{\omega(AA^\dagger)}\omega(AXA^\dagger)$ $(X\in\M)\in E(\M)$. Notice that, independently of the initial state $\omega$, $\omega_A$ is normalized: $\omega_A(\id)=1$.

Suppose now that $\omega_0\in E(\M)$ exists such that $\omega_0(XC)=0$, where $C\in \M$ satisfies the following condition: $C\,C^\dagger-C^\dagger\,C=\id$. In a quantum mechanical language, we would call $C$ and $C^\dagger$ {\em bosonic operators}. We further assume that $\omega_0(\id)=1$. In case it is not, we can normalize the state as shown before, still remaining in $E(\M)$. If we now introduce the operator $\hat N=C^\dagger C$, we can prove the following:
\begin{prop}\label{prop51}
	{Let us define $\omega_l(X)=\frac{1}{\omega_{l-1}(CC^\dagger)}\omega_{l-1}(CXC^\dagger)$, $X\in\M$ and $l\in\mathbb{N}$.
	Then $\omega_l\in E(\M)$ and $\omega_l(X\hat N)=l\omega_l(X)$, $X\in\M$ and $l\geqq 0$,
	that is, $\omega_l$ is an eigenstate of $\hat N$ with eigenvalue $l$.
	}
\end{prop}
\begin{proof}
	First of all we need to check that $\omega_{l}(CC^\dagger)\neq0$ for all $l\geq0$, to make sure the definition of $\omega_l(X)$ makes sense. We first observe that  $\omega_{0}(CC^\dagger)=\omega_0(\id+C^\dagger C)=\omega_0(\id)+\omega_0(C^\dagger C)=1$, because of the above properties of $\omega_0$. Now, let us suppose that $\omega_{l}(CC^\dagger)\neq0$ for a given $l$. Of course, this means that $\omega_{l}(CC^\dagger)>0$. We want to check that $\omega_{l+1}(CC^\dagger)>0$ as well. Indeed we have
	$$
	\omega_{l+1}(CC^\dagger)=\frac{1}{\omega_{l}(CC^\dagger)}\omega_{l}(CCC^\dagger C^\dagger)=\frac{1}{\omega_{l}(CC^\dagger)}\omega_{l}(C(\id+C^\dagger C)C^\dagger)=$$
	$$=1+\frac{\omega_{l}((CC^\dagger)(CC^\dagger)^\dagger)}{\omega_{l}(CC^\dagger)}\geq1,
	$$
	clearly. Next we observe that each $\omega_l\in E(\M)$ since, as we have observed, $E(\M)$ is closed under the operation mapping $\omega_{l-1}$ to $\omega_l$. 
	
	To prove now that $\omega_l$ is an eigenstate of $\hat n$ with eigenvalue $l$ we use induction on $l$. For this we observe that $\omega_0(X\hat N)=\omega_0((XC^*\dagger)C)=0$, in view of the definition of $\omega_0$.
	
	Next, let us assume that $\omega_l(X\hat N)=l\omega_l(X)$, $\forall X\in\M$. We need to check that $\omega_{l+1}(X\hat N)=(l+1)\omega_{l+1}(X)$, $\forall X\in\M$, as well. Indeed we have
\begin{eqnarray*}
\omega_{l+1}(X\hat N)
&=& \frac{1}{\omega_l(CC^\dagger)}\omega_l(CXC^\dagger CC^\dagger)\\
&=& \frac{1}{\omega_l(CC^\dagger)}\omega_l(CXC^\dagger(\id+C^\dagger C))\\
&=& \frac{1}{\omega_l(CC^\dagger)} (\omega_l (CX C^\dagger)+\omega_l(CXC^\dagger C^\dagger C))\\
&=& \frac{1}{\omega_l(CC^\dagger)} \omega_l(CX C^\dagger)+\frac{1}{\omega_l(C C^\dagger)}(l\omega_l(CXC^\dagger))\\
&=& \omega_{l+1}(X)+l\omega_{l+1}(X)\\
&=& (l+1)\omega_{l+1}(X),
\end{eqnarray*}
for all $X\in\M$.
	
	\end{proof}

A similar result can be obtained for a more general situation, for the so-called {\em abstract ladder operators}, \cite{bagspringer,bag2024}.

\begin{prop}
	Let us consider two elements $H_0=H_0^\dagger, Z\in\M$, such that $H_0Z-ZH_0=\alpha Z$, for some $\alpha\in\mathbb{C}$. Suppose that there exists an eigenstate $\nu_0$ of $H_0$ with eigenvalue $\gamma_0$
	and define $\nu_l=\frac{1}{\nu_{l-1}(Z^\dagger Z)}\nu_{l-1}(Z^\dagger XZ)$ for all $l\geqq 1$ and $X\in\M$.
	Then, if $\nu_{l}(Z^\dagger Z)\neq0$ for all $l$, $\nu_l$ is an eigenstate of $H_0$ with eigenvalue $\gamma_0+l\alpha$ for all $l\geqq0$.
\end{prop}


We will not give the proof of this claim, since it is similar to the proof of Proposition \ref{prop51}. These two propositions describe natural conditions which produce, out of a fixed element in $E(\M)$, other elements of the same sets all being eigenvectors of certain elements of $\M$ which have a crucial relevance in physics: the {\em number operator} $\hat N$, and the {\em Hamiltonian} $H_0$. Notice that, in both these cases, the eigenvalues of the operators considered are real. This is not a big surprise, since we have already seen that the eigenvalues of a self-adjoint operator are real. 

%

%
%
%
%
%

{ It is natural now to consider this problem: if we have two eigenstates of some given self-adjoint operator, with two different eigenvalues, is it possible, and in which sense, to deduce that the two eigenstates are orthogonal. In \cite{bit} we have considered a similar problem from the point of view of the GNS representation in a very specific case, where an unique Hilbert space for different sesquilinear forms can be defined. Something similar could be repeated here, but we will not consider this aspect of the construction here, since there are not many differences with respect what we have discussed in \cite{bit}. It would be interesting to set up a more general approach, on the same lines as in \cite{denittis}, but in the O*-algebraic settings many delicate aspects appear, which should be considered properly. We postpone this analysis to a future paper.

\bibliographystyle{amsplain}

\end{document}